# Investigation of Arrott plot and magnetocaloric effect in the complex $CaMn_7O_{12}$ perovskite

Parul Jain[1], Babita Ingale[1], Gyaneshwar Sharma[2], L.C. Gupta[3], A.K. Ganguli[3,4], Ratnamala Chatterjee[1,*]

[1]*Magnetics and Advanced Ceramics Laboratory, Department of Physics, Indian Institute of Technology Delhi, Hauz Khas, New Delhi-110016, India*

[2]*Department of Physical Science, Indian Institute of Science Education and Research, Knowledge city, Sector 81, SAS Nagar, Manauli PO 140306, Mohali, Punjab, India*

[3]*Department of Chemistry, Indian Institute of Technology, New Delhi 110016, India*

[4]*Institute of Nano Science and Technology, Habitat Centre, Mohali 160062, India*

## Abstract:

Detailed magnetic studies including magneto-caloric measurements on magnetic multiferroic quadruple perovskite $CaMn_7O_{12}$ are presented. Based on the collective response of Arrott plots and $\Delta S_M$ (T), a magnetic phase diagram of $CaMn_7O_{12}$ is suggested. A new magnetic transition at $T_{N3}$ ~20 K where the system changes from noncollinear AFM to collinear AFM is reported. An anomaly observed in both $\Delta S_M$ (T) and Arrott plots around 64 K has been attributed to high external field induced spin canting leading to change in magnetic order inducing phase transition. Magneto-caloric effect in this material is presented for the first time. The peak value of change in isothermal magnetic entropy ($-\Delta S_M$) is 1.3 J/K-Kg and the value of refrigeration capacity is reported to be 34.5 J/Kg for the field of 7 T.

## I. INTRODUCTION

Magnetoelectric multiferroic materials having (anti)ferromagnetism and ferroelectricity in the same phase have gained considerable interest due to their fundamental and perceived technological applications[1-6]. These multiferroic materials allow ferroelectric polarizations to be controlled by tuning their magnetism, owing to their strong intrinsic magnetoelectric couplings. Among different multiferroic materials[7-8] with various crystal structures, some perovskite manganites with ferroelectricity driven by magnetic orders are of particular interest. $CaMn_7O_{12}$, a member of quadruple perovskite family of manganites has recently been reported to show large ferroelectric polarization below its helical magnetic ordering temperature $T_{N1}$ ~90 K[5]. The compound is mixed valent with the $Mn^{3+}$ to $Mn^{4+}$ ratio 6:1 and remains in charge-ordered state below 250 K. Further, another phase transition which is also magnetic in nature has been reported[5,9] at $T_{N2}$ = 45 K in this compound. The measured polarization (2870 C/m$^2$) is the highest[5] measured magnetically induced ferroelectric polarization, persisting up to its Néel temperature $T_{N1}$ of 90 K. This discovery represents an important development in the field of magnetic ferroelectrics, as large polarization is crucial for electric manipulations through spins. It has been noticed that the phenomenon of breaking of inversion of spatial symmetry in $CaMn_7O_{12}$, leading to large polarization, is totally driven by magnetic ordering. Johnson *et al*[9] have explained the magnetic structure of $CaMn_7O_{12}$ below $T_{N1}$ by neutron powder diffraction experiments. They report that in this compound a helical magnetic structure with incommensurate propagation vector emerges in the temperature range $T_{N2} \leq T \leq T_{N1}$ indicative of a long range magnetic ordering. They also found that below $T_{N2}$ two propagation vectors were needed to index all the magnetic reflections.

In $CaMn_7O_{12}$ three different magnetic sublattices have been reported.[10] Spin orientations of magnetic structure in this composition depend on magnetic anisotropy of these three different magnetic sublattices. It is a well known[11] fact that the coupling of magnetic sublattices with external magnetic field leads to the simultaneous changes in magnetic part of total entropy; this makes $CaMn_7O_{12}$ an interesting candidate for studying Magnetocaloric effect (MCE).

Overall, earlier reports on $CaMn_7O_{12}$[10,12] suggest complex magnetic ordering at lower temperatures and the magnetocaloric data on this material is not yet reported. This motivated us to investigate the magnetic behavior in this quadruple perovskite in the low temperature region, specifically below 45 K. Further studies of magnetic states in is material are necessary[13] to clarify the origin of its remarkably large electric polarization.

In this work, we present the results of our detailed investigations of magnetic properties of $CaMn_7O_{12}$ in the range 2 K $\leq T \leq$ 90 K. Detailed temperature and field dependent magnetization behavior are analyzed using Arrott plot. Generally, Arrott plots have been widely used for determining Curie/Neel[14,15] temperatures in magnetic materials and for better understanding of magnetic interactions[16] in several magnetic systems including multiferroics.[17] The important results we report here are (i) a new magnetic transition temperature at $T_{N3}$ ~20 K is identified, in both Arrott plot as well as isothermal magnetic entropy curves, (ii) in the isothermal magnetic entropy plots a magnetic field induced anomaly is reported around $T_S$ = 64 K, which is also noticeable in Arrott plots; and last but not the least, (iii) based on the results obtained, a magnetic phase diagram for $CaMn_7O_{12}$ is suggested.

## II. RESULTS:

The double perovskite system $CaMn_7O_{12}$ is synthesized by wet sol gel method. $Ca(NO_3)_2 \cdot 4H_2O$ and $Mn(NO_3)_2$ (Merck, 99.9% purity) were mixed in their stoichiometric ratio with equal amount of de-ionized water and mixed well on hot plate magnetic stirrer at 70˚C for one hour. Then ethylene glycol is mixed to the solution and resulting solution is heated at 110˚C till the solid resin is obtained. This resin is kept in oven for 2 hours at 250˚C for complete combustion. Obtained powder is calcined in three steps 1)800˚C, 2)925˚C, and 3)950˚C for 48 hours each with intermediate grinding to avoid any $Mn_2O_3$, $Mn_3O_4$, and $CaMnO_3$ impurities[12]. This calcined powder is mixed with PVA 4 % by weight and then pressed into pellet and sintered at 965˚C for 75 hours. Powder X-ray diffraction (XRD) pattern obtained using Panalytical X-ray diffractometer with filtered Cu-Kα, wavelength (λ = 1.5418 Å) shown in Fig. 1, confirms[18] the formation of a single phase $CaMn_7O_{12}$. The XRD pattern could be fitted using Lebail fitting for the $R\bar{3}$ space group (as shown in Fig. 1 (b)). Lattice parameters found are $a = b = 10.44$ Å and $c = 6.34$ Å with pseudo-cubic angle = 90.33˚; are in good agreement with the literature values[17] Magnetization versus temperature (*M* versus *T*) at different fields (500Oe $\leq H \leq$ 5T) and Magnetization versus field (*M* versus *H*) measurements in the temperatures range (2K $\leq T \leq$ 300K) were performed using Quantum Design MPMSXL-7 SQUID magnetometer. For *M* versus *T* measurements, the sample was first cooled in zero-field down to the lowest temperatures (2 K), then the desired fields were applied and data were collected during heating. This is denoted as zero- field cooled (ZFC) curve. After the sample reached ~ 200 K (paramagnetic region), data were recorded during cooling down process to the lowest temperature of 2 K (field cooled cooling or FCC) and once again the data were collected as the temperature was increased from 2 K to 200 K (field cooled warming or FCW).

## A. Magnetization versus Temperature:

Figure 2 shows the *M* versus *T* curves measured at 500 Oe. As expected, a small kink in *M* versus *T* at $T_{N1}$ = 90 K (upper inset (a) of figure 2 magnifies the area to clearly show this kink) is noted, indicating a magnetic phase transition. On further cooling down to $T_{N2}$= 45 K, ZFC and FCC curves diverge from each other and a hysteresis is observed in FCC and FCW in the temperature range 15 K ≤*T*≤ 45 K. The existence of thermal hysteresis below 45 K clearly indicates the first order nature of the magnetic transition in this range. Lower inset of figure 2 shows 1/$\chi$ versus temperature (1/$\chi$-*T*) curve confirming the $T_{N1}$ transition at 90 K. A paramagnetic (PM) behavior is observed above $T_{N1}$ (see Curie-Weiss law fitting in inset (b) of Fig. 2). The extrapolated Curie-Weiss temperature is ~ −50 K. Such a Curie-Weiss behavior and a negative intercept confirm the PM to AFM transition at $T_{N1}$.

## B. Magnetization versus field:

The Magnetization (*M*) versus Field (*H*) data for the sample at different temperatures in the range 2K ≤ T ≤110 K are shown in Fig 3. Curves below $T_{N2}$ ~45 K show magnetic hysteresis and the coercivity ($H_C$) is seen to increase with decreasing temperature, as shown in the inset of fig 3.

## C. Isothermal magnetic entropy:

The virgin curves for all temperatures in the range 2 K ≤ T ≤ 100 K are shown in figure 4. From Fig. 4, $\Delta S_M(T)$ is estimated in this temperature range using Maxwell's equation,

$$\Delta S_M = \int_0^{H_{Max}} \left(\frac{dM}{dT}\right) dH \qquad \text{----- (1)}$$

The thermal profiles of change in isothermal magnetic entropy are then plotted as shown in Fig. 5. A few prominent features of Fig. 5 are:

(i) As strength of magnetic field is increased in isothermal manner, magnetic entropy of the system is decreased and the system becomes more ordered, (ii) the magnetic transitions labeled with $T_{N1}$ and $T_{N2}$ show significant changes in $-\Delta S_M(T)$ plots, (iii) interestingly, an inverse magnetocaloric effect (inverse MCE) is observed below 20 K suggesting a new magnetic transition at $T_{N3}$ ~20 K and last but not the least (iv) a field dependent anomaly visible for higher fields only (H ≥ 2T) in $-\Delta S_M(T)$ plots is observed at 64 K.

## D. Magnetization and Arrott plot:

Arrott plot is an alternative means of analysis, which is both more direct and physically more transparent. In order to investigate critical behavior of spins in the vicinity of transition temperatures, Arrott plots for the sample are shown in figure 6(a). The features of interest in this figure can be summarized as follows:- (i) Above 90 K ($T_{N1}$), curves are straight upwards showing paramagnetic behavior, (ii) below 90 K all the curves show a curvature: (a) 45 K ≤ $T$ ≤ 90 K, concave curvatures and (b) for $T$ < 45 K convex curvatures were noted, (iii) with increasing temperatures in the range 2 K ≤ $T$ ≤ 20 K, the plots are displaced to the left and then for $T$ > 20 K as the temperature is increased, plots start getting displaced to the right shown clearly in fig 6(b), (iv) interestingly, this anomalous behavior below 20 K is observed for low fields only; whereas, at higher fields Arrott plots displace continuously towards right with increasing temperatures, (v) as highlighted in the inset of Fig. 6(a), at high fields, a noticeable kink around $T$ = 64 K curve is clearly seen.

## III. DISCUSSION:

We are now in a position to discuss the collective response of $CaMn_7O_{12}$ samples as analyzed by Arrott plots and magnetocaloric measurements in the range 2 K ≤ $T$ ≤ 90 K. Above 90 K, only paramagnetism exists. In the range 20 K ≤ $T$ ≤ 90 K, with increasing magnetic field magnetic entropy decreases i.e. the system becomes more ordered. However, below 20 K magnetic entropy increases with increasing magnetic field i.e. the disorder increases in the system showing inverse MCE. Such a change in magnetic entropy from negative to positive value has been often observed in systems displaying first-order magnetic transitions such as AFM/FM[19], AFM/FI (ferrimagnetic)[20] or collinear AFM/non collinear AFM [21-22].

## A. Magnetic behavior in the range 20 K < $T$ ≤ 90 K

As reported in literature, in $CaMn_7O_{12}$, below $T_{N1}$ (~ 90 K), the spins within the sublattices start coupling with each other and acquire non collinear spin helical structure[10]. It is also known that in general, the isothermal exposure of non-collinearly ordered antiferromagnets to magnetic field results in a decrease in magnetic entropy[23]. It can be noted that we too find a decrease in magnetic entropy of $CaMn_7O_{12}$ (i.e., increase in -$\Delta S_M(T)$) in the temperature range 20 K < $T$ ≤ 90 K, on its isothermal exposure to magnetic field (see Fig 5). This clearly indicates that our magnetocaloric measurements validate the non-collinear antiferromagnetic behavior of $CaMn_7O_{12}$ in the temperature range 20 K < $T$ ≤ 90 K. The broad peaks in -$\Delta S_M(T)$ plots, which become more pronounced as the field is increased, is perceived to be due to the spin helical ordering.

As has been evidenced by neutron diffraction studies[9], heat capacity[12] and dielectric measurements[24], there exist two different transitions at 45 K and 90 K in this compound. Accordingly, clear changes in magnetic entropy curves (Fig. 5) at the above two specific temperatures are noted in -$\Delta S_M(T)$ plots. The change of curvatures at 45 K and 90 K in Arrott plots are also indicative of magnetic transition at these two temperatures.

As we take a closer look at these two temperatures in various magnetic data, we note from the inset of Fig.3 that at/around $T_{N2}$ (~45 K) appearance of coercivity in M-H data is clearly seen. The coercivity values increase with decreasing temperature; indicating a possible starting of appearance of weak ferromagnetism below $T_{N2}$ in this compound. A closer inspection of field dependent -$\Delta S_M$ behavior at two magnetic transitions, $T_{N1}$ and $T_{N2}$, reveal that, although at both $T_{N1}$ and $T_{N2}$, we see a dip in -$\Delta S_M$, there is a distinct difference in behavior of -$\Delta S_M(T)$ plots at these two temperatures- (a) the dip in -$\Delta S_M$ at $T_{N1}$ (~90 K) increases with increasing field, whereas (b) at around $T_{N2}$ (~ 45 K), the dip decreases with increasing field. This difference in behavior could be coming from different degree of spin canted magnetic structure in $CaMn_7O_{12}$.

At $T_{N1}$, the helimagnetic structure is due to a strong correlation among nearest neighbor spins[9], hence, the dip increases with increasing magnetic field. Whereas, around $T_{N2}$, the ferromagnetic contribution (as discussed above), that increases with increasing magnetic field, possibly leads to a decrease in the dip. Near magnetic transition $T_{N2}$, a short range ferromagnetic order and long range antiferromagnetism compete with each other. The presence of competing magnetic interactions at $T_{N2}$ = 45 K is also indicated in Arrott plots. A kink (at 10 kOe) in the 45 K curve in fig 6(a) clearly indicates a combination of concave and convex curvatures.

An anomaly around 64 K is observed in Arrott plots for higher fields (shown in the inset of Fig. 6(a)); which is in support with the anomaly clearly observed in -$\Delta S_M$ (*T*) plots. To have a better insight around this temperature we also performed *M-T* measurements in varying fields, 0.5 T ≤ H ≤ 5T. A field induced hysteresis around this temperature at higher fields of 4 T and 5 T (shown in Fig. 7 (a)), supports our isothermal magnetic entropy data. A field induced magnetic structure emerges at $T_S$ = 64 K; as is reported in literature of various materials showing first order phase transformation[27].

## B. Coupling of magnetic moments below 20 K ($T_{N3}$):

The thermal profile of -$\Delta S_M$ below 20 K, suggests an additional magnetic activity/behavior by exhibiting inverse MCE. Inverse MCE has been often observed in systems displaying transitions such as AFM/FM, AFM/FI or collinear AFM/non collinear AFM. In our earlier work[15] we have demonstrated that although usually the Arrott plots are used to determine the $T_C$ in weakly ferromagnetic materials, they can also be used to estimate the $T_N$ in materials with multiple magnetic interactions[15]. As it can be observed in Fig. 6(b), the Arrott plots in the range 2 K < *T* ≤ 20 K are displaced to the left as the temperature is increased; whereas, after 20 K the plots start displacing towards right with increasing temperatures. This behavior in Arrott plots also reinforces that 20 K is an antiferromagnetic ordering temperature. However, it should be noted that this behavior below 20 K is observed for low fields only; at higher fields Arrott plots displace continuously towards right with increasing temperature. Due to the strong coupling between ferromagnetic and antiferromagnetic moments at lower fields, the antiferromagnetic order renormalizes the ferromagnetic order in such a way that the increase in the ferromagnetic moment with field is reduced in comparison to the case where there is no coupled antiferromagnetic order. As a result of this renormalization the antiferromagnetic moments dominate for lower field strengths. However, the antiferromagnetic state becomes unstable for large fields and the system switches from a predominantly antiferromagnetic state to a ferromagnetic state for higher fields, in this tempetarure range (2 K $< T \leq 20$ K).

The appearance of this inverse MCE in $CaMn_7O_{12}$ suggests that below $T_{N3}$, spin ordering may be collinear antiferromagnetic. From the results of $M$ vs. $H$ curves (Fig. 3), magnetic entropy (Fig. 5) and from Arrott plots (Fig. 6(a)) one can interpret that the $CaMn_7O_{12}$ system might be changing from noncollinear AFM to collinear AFM with increased ferromagnetic orders at lower temperatures (T $\leq$ 20 K); or there is a kind of collinear antiferromagnetic ordering with ferromagnetic planes in the system as reported for $Ni_3TiO_6$.[26]

## C. Magnetic phase diagram:

On the basis of magnetization measurements it has been already inferred that $CaMn_7O_{12}$ remain in paramagnetic state above 90 K, below which it gets antiferromagnetically ordered. At around 45 K and 20 K again magnetic phase transitions occur due to complex magnetic ordering. In order to elaborate these magnetic transitions, we have plotted $\chi \times T$ as a function of temperature as shown in Fig. 7(b). Based on our results, the $\chi \times T$ vs.$T$ can be divided in different regions – PM (paramagnetism), AFM-I, AFM-II and AFM-III. A comparison of anomalies corresponding to transition $T_{N1}$ and $T_{N2}$ (discussed above in Section IIIA) are suggesting that these are not proper AFM collinear magnetic ordering. The canted symmetry is strengthened on lowering thermal energy. Sharp upturn in $\chi \times T$ in AFM-II region corresponds to the highly frustrated magnetic symmetry or superparamagnetic clusters in the system. A new region AFM III has been added below 20 K on the basis of magneto caloric as well as Arrott plot analyses. Below $T_{N3}$ the system shows collinear antiferromagnetic ordering along with the strong coupling between ferro and antiferro moments. A new division within AFM-I at $T_S$ temperature is suggested on the basis of isothermal magnetic curves. A field induced magnetic structure emerges at this temperature inducing a phase transition.

Figure 8 shows magnetic phase diagram of $CaMn_7O_{12}$, derived from the collective response of Arrott plot and $\Delta S_M$ (T). The splitting of magnetic region is based on slope variation of Arrott plot and $\Delta S_M$ (T) plot. The magnetic region labeled as PM denotes paramagnetic region. AFM-Heli region, below 90K, represents incommensurate helical spin ordering. In this region linearity of Arrott plot is modified with the magnitude of external magnetic field. In low fields, the Arrott lines exhibits negative slope (region AFM-HC(I)), suggests that internal magnetic field are dominant in nature. Under high field, same transition belongs to the second order (labeled by AFM-HC(II)). Below 64K, field mediated magnetic transition (AFM) is observed. In low field region this transition remains insensitive towards physical parameter magnetic field and temperature. Field with intermediate magnitude exhibits Arrott lines with moderated positive slope (AFM-HC(III)). Under high field, the Arrott lines possess enhanced slope. The evolution of novel magnetic phases, AFM-HC(II) and AFM-HC(III), are realized against the revision of spatial distribution of adjacent spin. This corresponds to some spin flop type activity, labelled with "spin flop". Below 45 K, Arrott lines exhibit different degree of sensitivity to external field. Low field and high field region are subdivided in AFM-NCL(I) and AFM-NCL(II) respectively. The Arrott lines shift closer to each other on lowering thermal energy. This suggests that chirality of helical spin structure is revised and canted feature of adjacent spins is lowered. The Arrott lines in AFM-CL region present non-collinear AFM and collinear AFM phase transition.

## Conclusion:

In summary, we have tried to explain the magnetic interactions in this quadruple perovskite magnetic multiferroic material $CaMn_7O_{12}$. A new magnetic transition ~20 K could be probed using magnetocaloric and Arrott plot analysis. Based on these results a magnetic phase diagram for $CaMn_7O_{12}$ in the entire temperature range of 2 K < $T$ ≤ 110 K is suggested. The peak value of -$\Delta S_M$ is reported to be~ 1.3 J/K-Kg at 51 K for 7 T field. Refrigerant capacity (RC) is a figure of merit of a magnetocaloric system and is calculated to be ~34.5 J/Kg for $CaMn_7O_{12}$ which is comparable to the other multiferroic materials like $YMnO_3$[28] and intermetallic alloys such as NiMnSn[29]. The continuous evolution of Arrott plot at low fields from the concave to convex curvature with decreasing temperature and unique behavior around new proposed $T_N$ is explained on the basis of interplay between spin-spin interactions and coupling between magnetic moments. An anomaly around 64 K has been observed in $\Delta S_M$ (T) profile as well as in Arrott plot at higher fields. This has been attributed to high external field induced spin canting leading to change in magnetic order inducing phase transition. These new results i.e. a new magnetic transition at $T_{N3}$ ~20 K as well as anomaly observed at 64 K may attract the attention of researchers for further investigations in this system.

## Acknowledgement:

The authors would like to acknowledge SQUID facility at IIT Delhi, India, for magnetic measurements. The authors P.J. and B.I. would also like to acknowledge CSIR (India) for providing financial support.

**Figures:**

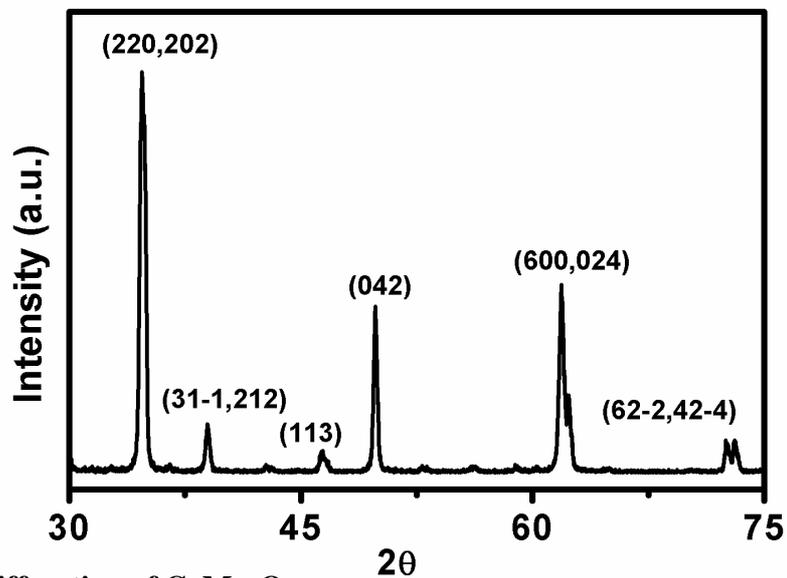

**Fig 1(a):** X-ray diffraction of $CaMn_7O_{12}$.

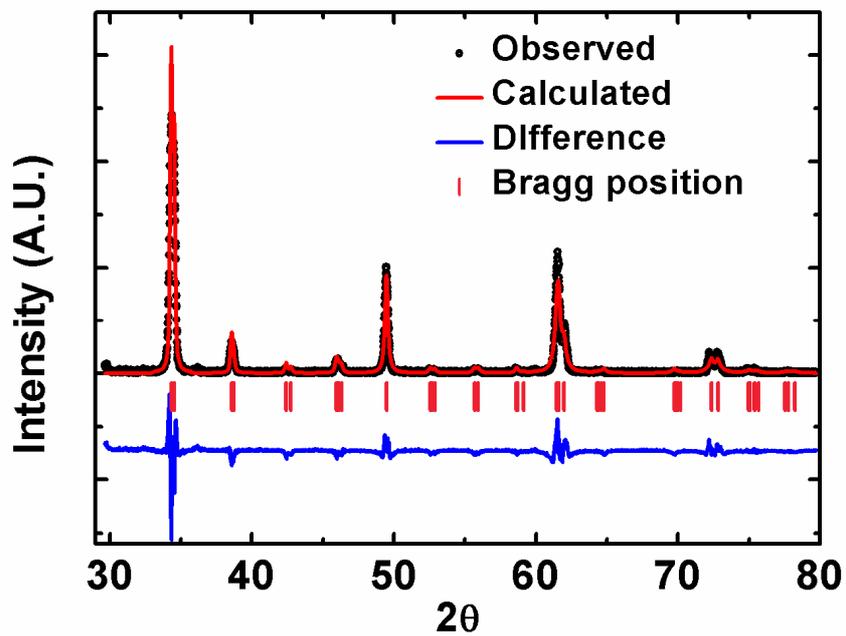

**Fig 1(b):** lebail fitted xrd pattern of $CaMn_7O_{12}$ at room temperature.

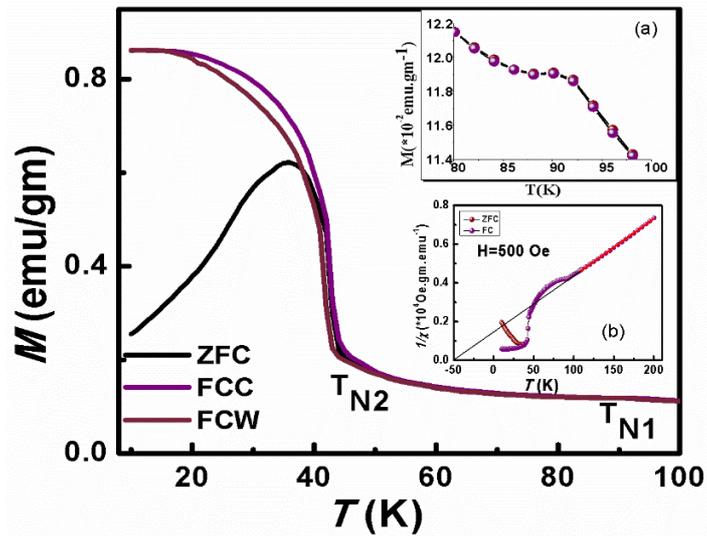

**Fig.2:** Magnetization (M) Versus Temperature (T) plots for $CaMn_7O_{12}$ in external magnetic field 500 Oe over the temperature range 10K-110K. Upper inset (a) shows the magnified M versus T curves for 500 Oe within low temperature range (80K ≤ T ≤ 100K). Lower inset (b) $\chi^{-1}$ versus Temperature (T) plots for $CaMn_7O_{12}$ in external magnetic field 500 Oe. Red line indicates the fitting of data using Curie-Weiss law.

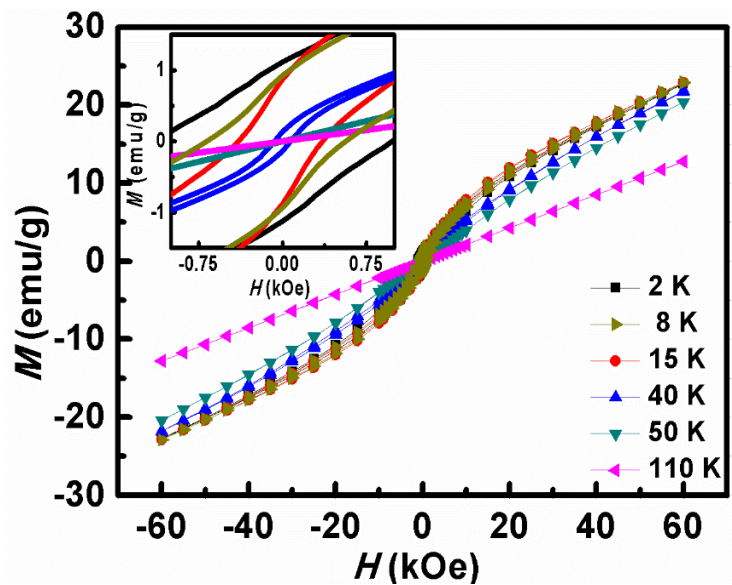

**Fig.3: Magnetization (M) versus field (H) curves for CaMn$_7$O$_{12}$ at different temperatures.**

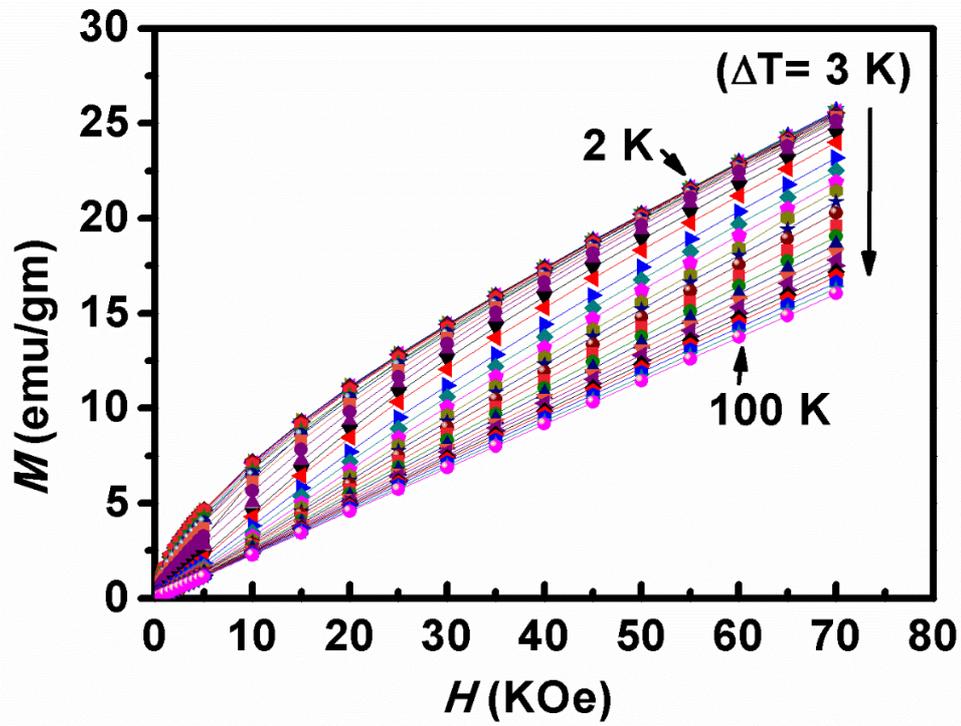

**Fig. 4: Isothermal magnetization (M) versus field (H) plots for sample in the temperature range 20 K to 100 K.**

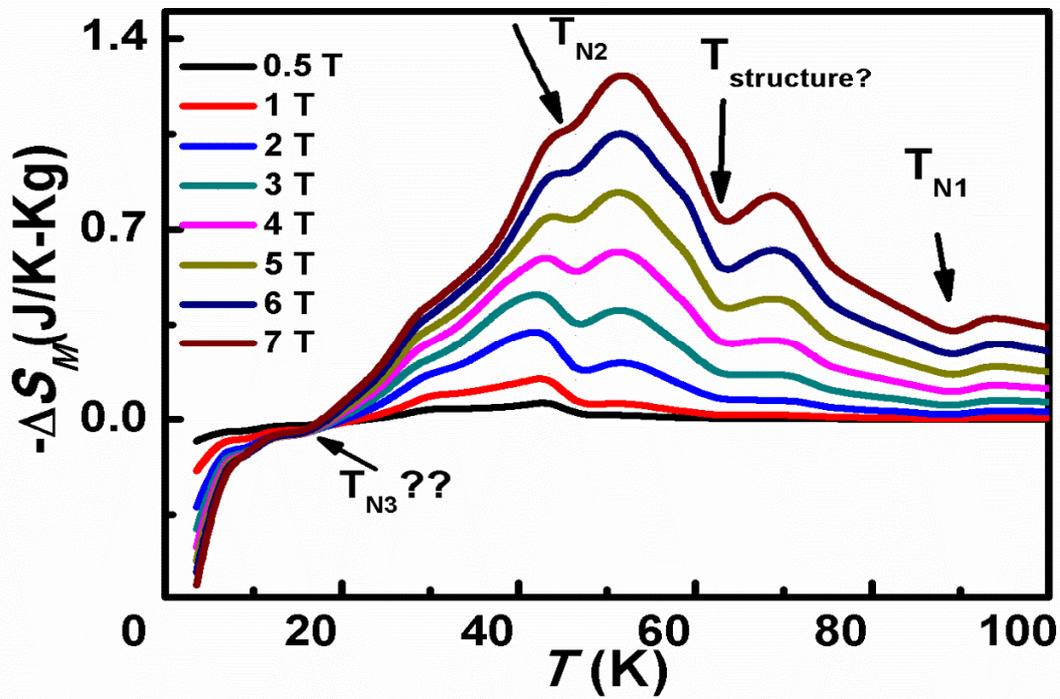

**Fig. 5:** Magnetic entropy change vs. temperature, -- $\Delta S_M$ (T), curves for temperature range 20 K- 110 K for constant magnetic fields from 0.005 -7 T.

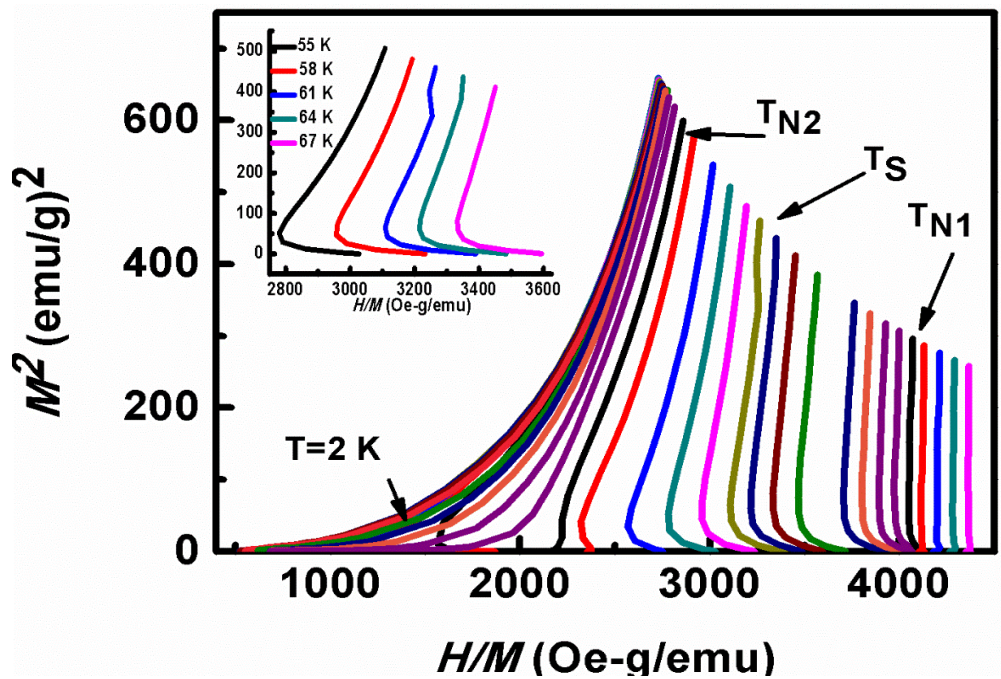

**Fig 6(a):** Arrott plot for $CaMn_7O_{12}$ over wide range of temperatures $2K \leq T \leq 100K$. Inset shows zoom view of Arrott plot for $CaMn_7O_{12}$ around 64 K.

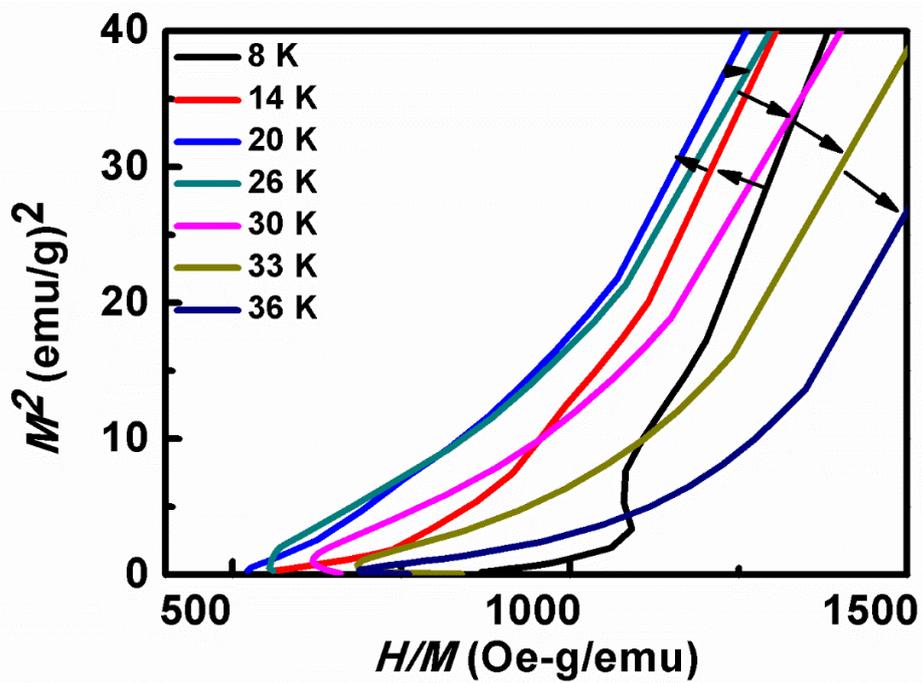

**Fig 6 (b):** Zoom view at lower temperature around $T_{N3}$.

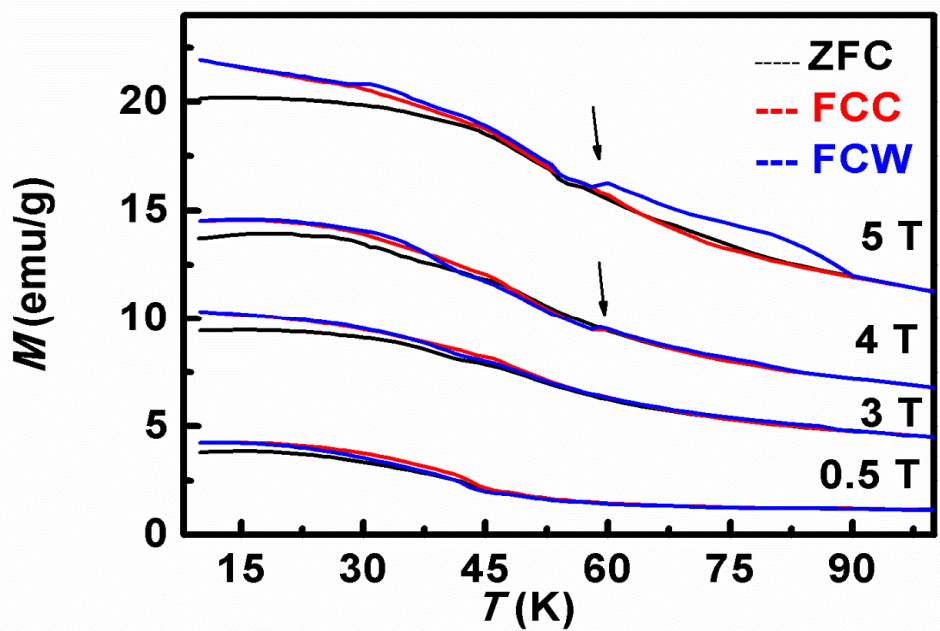

**Fig.7 (a): Magnetization (M) Versus Temperature (T) plots for $CaMn_7O_{12}$ in external magnetic field of 0.5 T, 3 T, 4 T and 5 T. over the temperature range 10K-110K.**

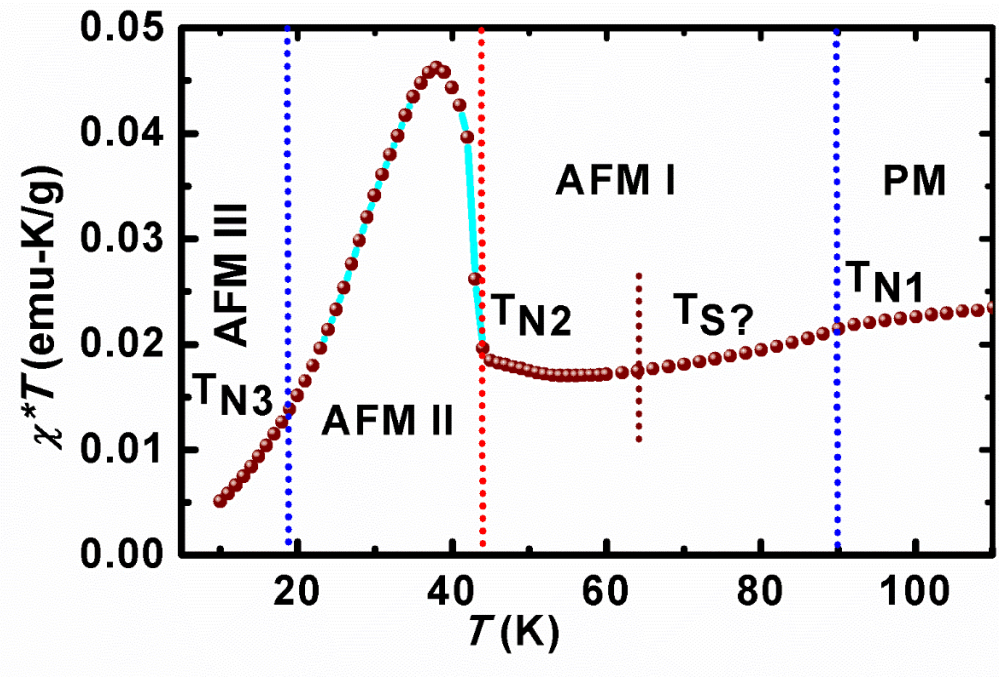

**Fig. 7(b): $\chi \times T$ versus temperature (T) curves for temperature range 2 K - 110 K for constant magnetic fields 500 Oe.**

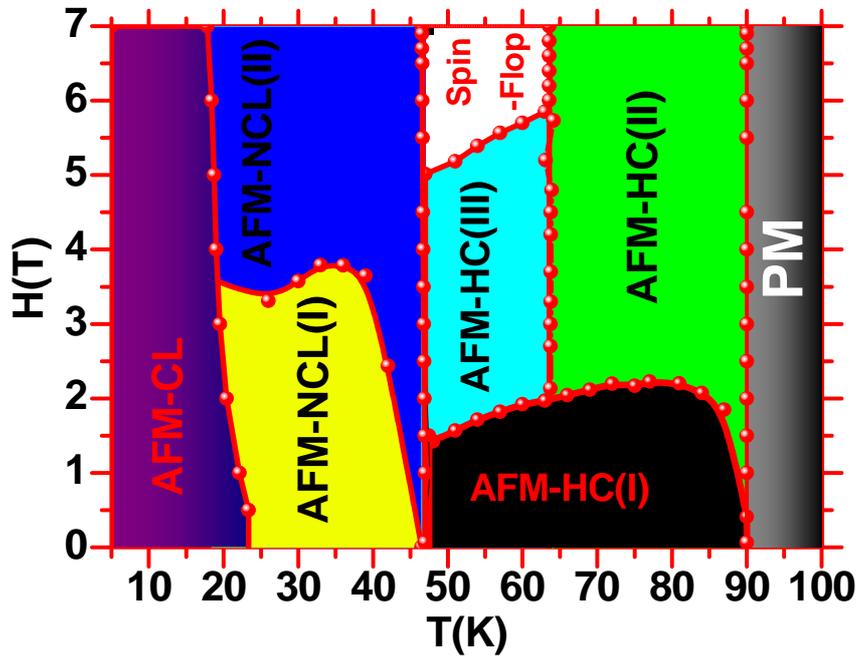

**H(T): Magnetic field in Tesla (y- axis), T(K): Temperature in Kelvin (x-axis)**

**AFM-CL: Antiferromagnetic-Collinear**

**AFM-NCL: Antiferromagnetic-Noncollinear**

**AFM-HC: Antiferromagnetic-Helicoidal structure**

**Fig. 8: Magnetic phase diagram of $CaMn_7O_{12}$ on the basis of Arrot plot and magnetic entropy $\Delta S_M$ (T) plot.**